# Transport and magnetic properties of $Fe_{1/3}VSe_2$

**C S Yadav and A K Rastogi**

School of Physical Sciences, Jawaharlal Nehru University, New Delhi-110067, India.

E-mail: shekharjnu@gmail.com, akr0700@mail.jnu.ac.in

**Abstract**

Electrical conductivity, thermopower and magnetic properties of Fe-intercalated $Fe_{1/3}VSe_2$ has been reported between 4.2K - 300K. We observe a first order transition in the resistivity of the sintered pellets around 160K on cooling. The electronic properties including the transitional hysteresis in the resistance anomaly (from 80K-160K) are found to be very sensitive to the structural details of the samples, which were prepared in different annealing conditions. The thermopower results on the sintered pellets are reported between 10K - 300K. The magnetic measurements between 2K - 300K and up to 14 Tesla field show the absence of any magnetic ordering in $Fe_{1/3}VSe_2$. The magnetic moment per Fe -atom at room temperature (between 1.4 $\mu_B$ – 1.7$\mu_B$) is much lower than in previously reported anti-ferromagnetic $FeV_2Se_4$. Furthermore, the Curie constant shows a rapid and continuous reduction and combined with the high field magnetization result at 2K suggests a rapid decrease in the paramagnetic moments on cooling to low temperatures and the absence of any magnetic order in $Fe_{1/3}VSe_2$ at low temperatures.

## Introduction

The structural, magnetic and optical properties of the complexes of the 3d-metal atom intercalated in the layered compounds of Nb- and Ta- dichalcogenide have been extensively reviewed [1,2,3]. In these compounds, the transition metal atoms form an ordered 2×2 and √3×√3 hexagonal superstructure for the 25% and 33% intercalation respectively [4]. The 3d-moments are localized, and show a large anisotropy of paramagnetic susceptibility, and give magnetic order at temperatures between 20-120K via the RKKY exchange coupling through the conduction electrons--- the ferromagnetic one for V, Cr , Mn and the anti-ferromagnetic order for Co and Ni atoms. For the iron intercalates compounds, diverse magnetic behavior depending on the host lattice and the atomic-ordering have been reported [5]. In a recent study, the ferromagnetic compound $Fe_{1/4}TaS_2$ ($T_C$ = 160K) showed extreme anisotropy and sharp switching of magnetization at 3.7 Tesla field and 2K temperature [6,7].

Ternary compounds $FeV_2S_4$, $FeV_2Se_4$ and $FeTi_2Se_4$ are known to show $Cr_3S_4$ type layered structure [8]. In these compounds also a cation deficient layer (50% Fe atoms) alternate with the filled layer containing V or Ti-atoms, same as in above mentioned compounds of Nb and Ta. The magnetic properties of these compounds show significantly reduced effective moment between 3.0 - 4.2$\mu_B$ per Fe atom, contrary to 4.9 $\mu_B$ as expected for high spin state of $Fe^{2+}$ [9,10]. But the anti-ferromagnetic ordering temperatures $T_N$ in these compounds at 131K, 94.5K and 134K respectively remain quite similar to that found in Nb- and Ta-compounds [9,10].

We have undertaken a detailed study of the structural and electronic properties $Fe_xVSe_2$ compositions with x varying between 0 - 0.33. We will report separately the properties of the compounds having Fe-concentration x < 0.33 [11,12]. Here we report the properties of $Fe_{1/3}VSe_2$ compound and the effect of intercalate ordering in different phases. These phases were obtained by varying annealing conditions of preparation. We have used low temperature and excess Se-pressure for their synthesis in order to avoid self-intercalation of V atoms in our compound.

**Sample preparation**

$Fe_{1/3}VSe_2$ was prepared from the elements by solid state chemical reaction route in two-step process. At first FeSe and $VSe_2$ (with 5% excess Se) were prepared by reacting Fe (99.98%), V (99.995%) and Se (99.95%) in proper molecular ratio at $600^0$C, inside the evacuated sealed quartz tubes. FeSe and $VSe_2$ were then thoroughly grounded and mixed in 1:3 molar ratio. This mixture thus contains excess Se over $Fe_{0.33}VSe_2$ composition. It was vacuum-sealed in separate evacuated quartz tubes for further reaction at different temperatures of $550^0$C and $700^0$C for 5 days. The excess amount of Selenium was released at the cold end of the sealed tubes. The reacted charges were grounded and pelletized at 5 ton pressure before sintering them at the chosen temperatures. These preparation conditions gave samples (i) FVS700: annealed at $700^0$C for 15 days, (ii) FVS550A: annealed at $550^0$C for 7 days, (iii) FVS550B: long annealed at $550^0$C for 15 days, and (iv) FVS850Q: annealed at $850^0$C for 5 days followed by air quenching.

**Structure**

The X-ray diffraction patterns of our $Fe_{1/3}VSe_2$ samples using $CuK_\alpha$ radiation are presented in Figure 1. For the comparison we also show a pattern of $VSe_2$ that was used for the preparation of our compounds. It can be seen that for FVS700 sample the pattern is very similar to the parent $VSe_2$, indicating that in this sample Fe is intercalated without any changes in the structure of host lattice. We could refine the structural parameter using GSAS-EXPGUI program for powder X-ray diffraction [13] and were able to ascertain the likely occupancy of the Fe atoms. This routine gave very good fit of the pattern for the choice of Fe atoms on the 1b (0 0 0.5) site positions in the space group $P\overline{3}m1$ (no. 164) of $VSe_2$ [14,15]. We thus confirm a random occupancy of the octahedral sites by 33% of Fe atoms in between the layers. The hexagonal lattice

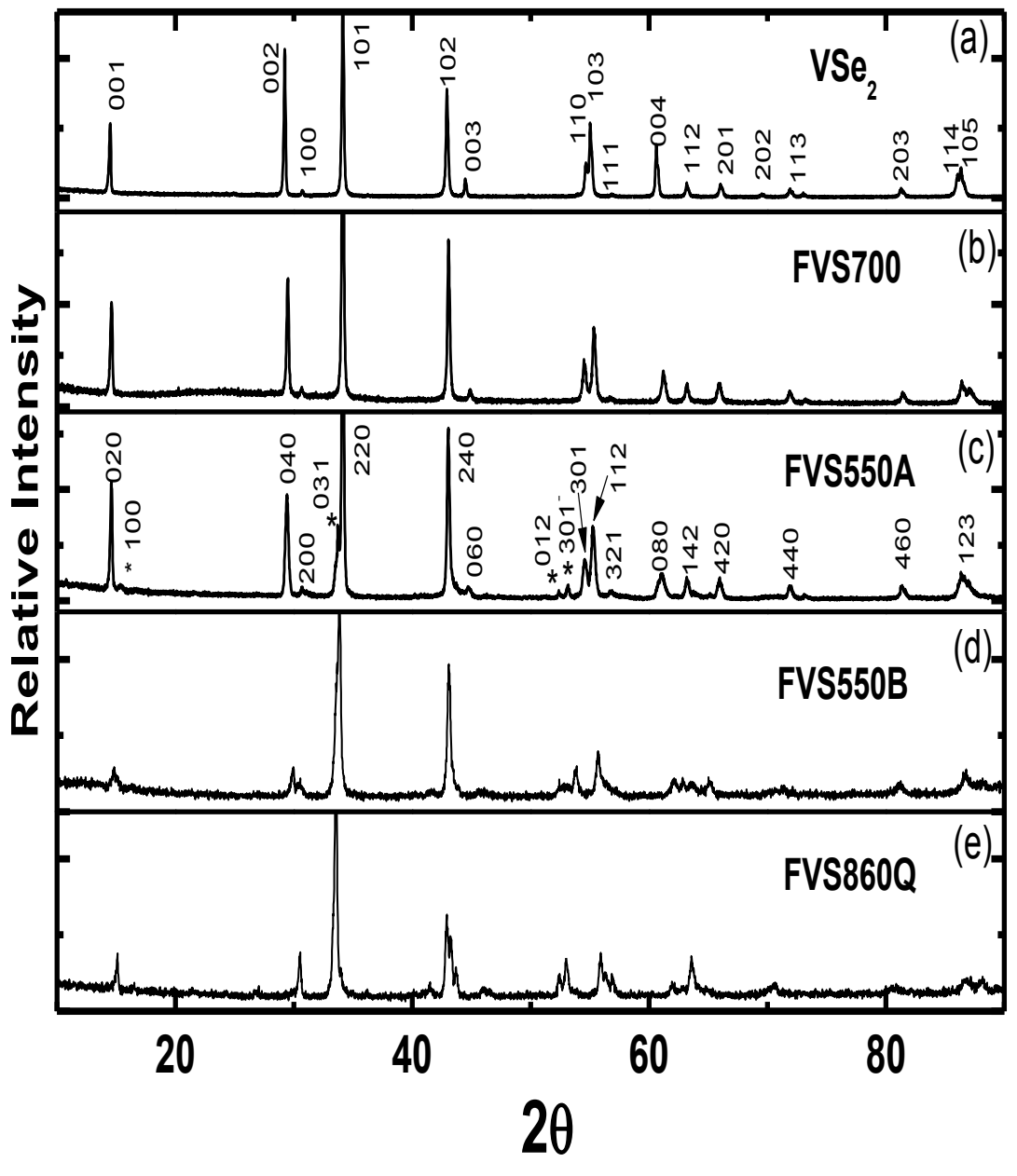

Figure1. powder X ray diffraction pattern of the $Fe_{0.33}VSe2$ compounds, (a) Parent compound 1T-VSe2, (b) FVS700, prepared at $700^0$C, (c) FVS550A, prepared at $550^0$C, (d) FVS550B, long annealed at $550^0$C, and (e) FVS850Q, phase quenched from $850^0$C.

parameters and the z- parameters of 2c (0 0 ±z) positions of the Se atoms obtained after refining the two structures are as follows;

$VSe_2$: a = 3.356Å, c = 6.104Å, and z = 0.2665, V = 59.53 $Å^3$, Z = 1.

$Fe_{1/3}VSe_2$: a = 3.365Å, c = 6.056Å, and z = 0.2537, V = 59.38 $Å^3$, Z = 1.

In 1T-$VSe_2$, the exceptionally large c/a ratio of 1.819 means a large trigonal elongation of the Se octahedra surrounding a V atom (the ideal c/a value is 1.633). Our results show that Fe intercalation reduces the c/a ratio considerably. Moreover, this reduction is primarily achieved from the decrease in the thickness (2cz) of the individual $VSe_2$ layers by the positive chemical pressure exerted by intercalated Fe atoms. This is a significant result and may explain the increase in the transition temperature from 110K in $VSe_2$ to a higher temperature of 160K in our Fe intercalated compounds due to similar reason as for the effect of the application of external pressure on $VSe_2$. This transition shift towards higher temperatures in presence of pressure is explained in terms of the pressure induced broadening of 3d-bands of Vanadium, which reduces the correlation effects among the electrons in a narrow band and enhances the CDW–instability [16].

*Super-lattice distortion*: For the compound FVS550A; prepared at $550^0$C, the peak positions and relative intensity of the pattern looks similar to the FVS700, except for the presence of some additional weak peaks ( marked by *) . A careful analysis revealed a super-lattice distortion of the hexagonal cell in this compound. We could index all the lines (fig 1c) using a monoclinic unit cell of the following dimensions;

FVS550A: a = √3×3.364Å, b = 3.527Å, c = 2×6.074Å and γ = $91.609^0$, V =249.6$Å^3$, Z= 4

The monoclinic cell dimensions for our $Fe_{1/3}VSe_2$ (FVS550A) compound can be compared with that reported for anti-ferromagnetic $FeV_2Se_4$ (a = √3×3.573Å, b = 3.47Å, c = 2×5.895Å and β = $91.60^0$, V = 253.14$Å^3$, Z = 4) [10]. Although both have similar structure and the size of the cell is also same, we found a different nature of distortion and the orientation of the supercell in our compound. These differences in the detailed structural features (yet unknown) of Fe sublattice of the two compounds may be the reason for the considerable differences in their magnetic properties.

We should notice that a rectangular supercell (with Z = 4) is possible only in case of the ordered arrangement of 1/4, 2/4 or 3/4 intercalate concentration. Hence in our $Fe_{1/3}VSe_2$ compound with 1/3 Fe-atoms, the ordering can only be short range. Moreover, the crystal will be subjected to the localized strains and the clustering of atoms by the accommodation of extra Fe atoms in the unit cell. These structural features will interrupt the long range ordering and are entirely consistent with the electrical and magnetic properties around a phase transition as reported below. The irreversible effects in the magnetization processes observed below 250K can be explained due to disordered freezing of the small number of magnetic clusters. In addition, the observation of a relatively broad hysteresis from 80K – 160K temperatures observed in the resistivity around structural transition reflects that the lattice is highly strained in FVS550A phase.

*Long annealing and quenching*: The X-ray pattern of the FVS550B, which is long annealed at $550^0$C for 15 days, showed significant broadening of the diffraction peaks and the absence of the lines corresponding to the *00l* reflections of the parent phase. This pattern permitted the estimation of approximate dimensions of the unit cell as follows;

FVS550B: a = √3×3.44, b = 3.45Å, c = 2× 5.97Å and γ = $91.74^0$, V= 245.77$Å^3$, Z=4.

The reason for the substantial change in the x-ray pattern after long annealing may be related to the releasing of elastic strains and the changes in the short range ordering of the supercell. We observe a near absence of (*00l*) reflections and the asymmetrical broadening of (*h0l*) peaks, indicating that the strains are released by introducing stacking disorder in our long annealed sample. The resulting domains are now better ordered as can be inferred from the sharp jump like changes in the resistance and the magnetic susceptibility at the transition reported below.

In fig 1e, we have shown the X-ray pattern of the FVS850Q phase, which is quenched from $850^0$C. Though it was not possible to calculate the parameters of the unit cell of this phase from the X-ray diffraction pattern, the pattern looks very much similar to the monoclinically distorted FVS550B along with additional splitting of some of the lines.

**Transport Properties**

*Resistivity*: We have shown the results of the measurement of the electrical resistance of the polycrystalline pellets of our compounds in figure 2. The resistivity of FVS700, FVS550A, FVS550B and FVS850Q at room temperature, were found to be 1.8mΩ-cm, 1.5 mΩ-cm, 1.3 mΩ-cm, and 0.5 mΩ-cm respectively. Except for the quenched phase FVS850Q, others show an increase in resistance on cooling. The large resistivity and its variation by less than 20%, in the temperature range 4.2K - 300K indicate that these compounds have a poor metallic conduction behavior. A clear transition like anomaly with the different extent of hysteresis can be seen on cooling in all of the phases

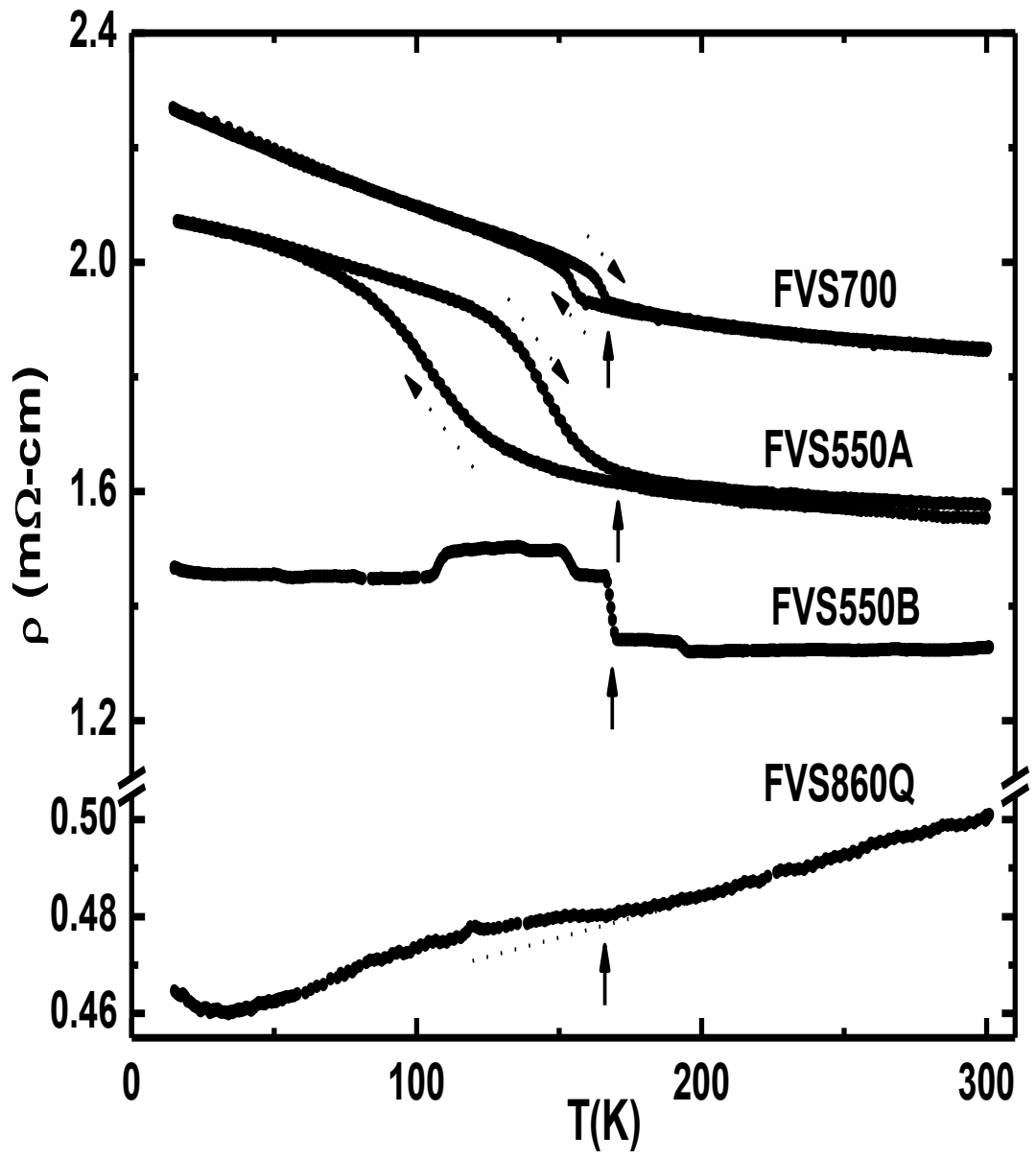

Figure2. The temperature dependence of Elctrical resistivity for the different phases of Fe0.33VSe2 showing the onset anomaly at 160K.

starting at 160K. In the quenched phase with metal like temperature dependence, the transition anomaly appears as a small bump between 80-160K break, which is quite similar to that found for 1T-$VSe_2$ under pressure [16]. A minimum in the resistance is also observed at 30K in this phase.

The largest hysteresis of the transition is found in the resistivity of FVS550A, in which a short range ordering of intercalated Fe atoms and related super-lattice distortions are found by X-ray analysis. As we have already mentioned above that this phase is highly strained due to the occupancy of extra Fe atoms on the vacancy ordered sites in $Fe_{1/3}VSe_2$. At the first order transition these local strains cause a broad hysteresis as observed in fig 2. We also notice that after the long annealing of this phase for 15 days, the resistivity changes with the sharp jumps in the temperature range 100K–160K. The long annealing removes the lattice strains in FVS550B and so the better ordered domains exhibit sharp transition.

*Thermoelectric Power*: The results for the Seebeck coefficient measurements between 15K – 300K on the sintered pellets of the FVS700 and FVS550A are shown in figure 3. The Seebeck coefficient is positive, and except for the saturation effect at high temperatures follows a metal like temperature dependence. The overall behavior of the thermopower is very similar to the other binary and ternary sulfides of Vanadium, viz. $V_{1+x}S_2$ and $Al_xVS_2$ [17], where the transport properties along with the variation of the Hall coefficient with temperature suggest a mixed conduction behavior due to the carriers in the overlapping chalcogen p–band and Vanadium 3d–band [12].

Our results on $Fe_xVSe_2$; x = 0 - 0.33 show that in general, the thermopower is relatively insensitive to the variations in intercalate concentration and also to the vacancy ordering related structural changes, specially when the measurements are performed on the respective polycrystalline pellets [12]. We have also found that the Seebeck coefficient of the polycrystalline $Fe_xVSe_2$ compounds remains same within 20% and shows same dependence on temperature for x = 0.02–0.33. Any visible effects in the thermopower related to the CDW-transition in these compounds were found to be absent [11,12].

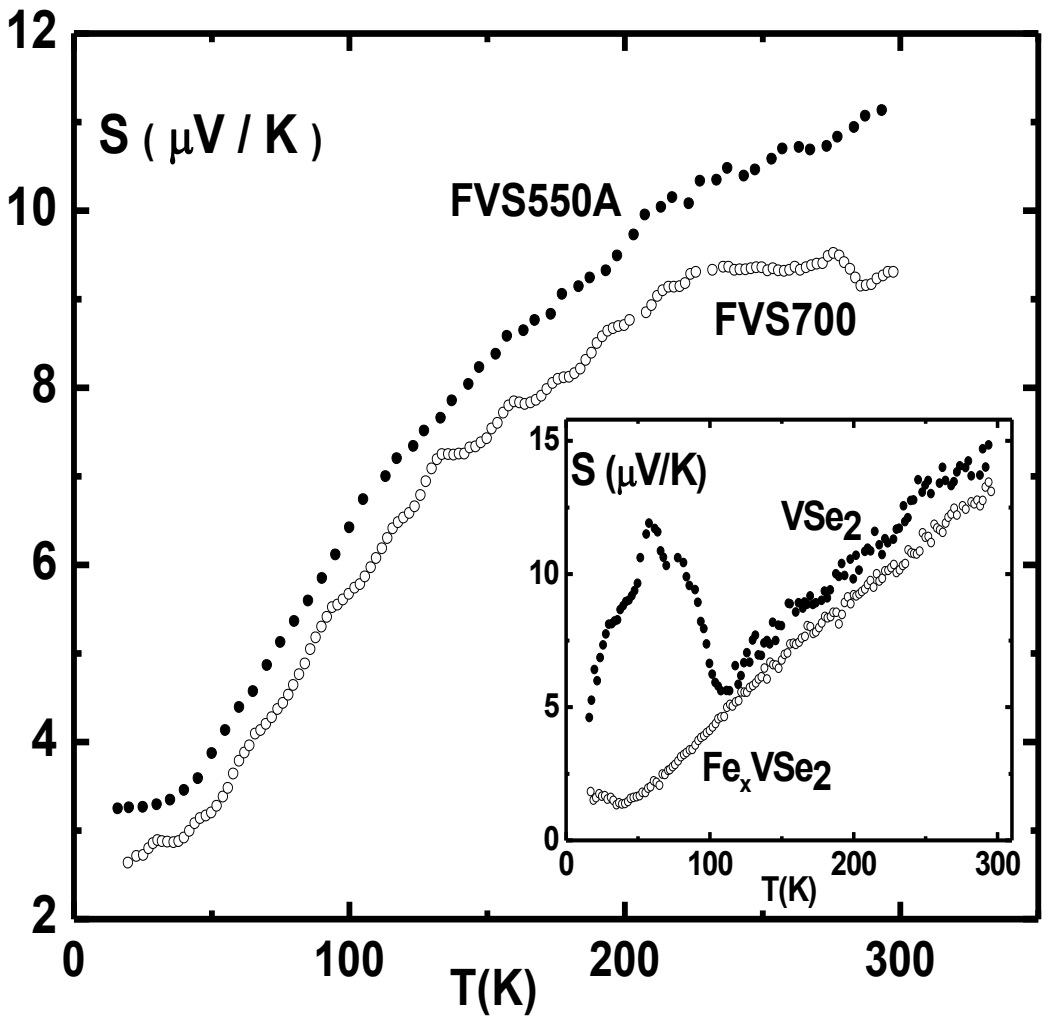

Figure 3. The Seebeck coefficient of the FVS550A and FVS700 phases, showing a saturating trend at higher temperatures. The Seebeck coefficient in the plane of $VSe_2$ and $Fe_xVSe_2$ are shown in the inset.

In contrast to the measurement on the polycrystalline samples, a very sharp jump like anomaly was measured in the Seebeck coefficient of a good quality single crystal flake of pure $1T\text{-}VSe_2$ at its 110K CDW transition, where the resistivity showed relatively weak feature. This jump is, however absent in the polycrystalline pellets of pure $1T\text{-}VSe_2$ (not shown here) and also in the flakes containing small but unknown quantity of Fe. We have plotted the Seebeck coefficient of the Flakes in the inset of Figure 3 for their comparison with the polycrystalline pellets of the present compounds. We can conclude that the grain boundaries in polycrystalline samples, as well as the crystal defects eliminate the sharp variations in the thermopower at the CDW transition observed in case of single crystal flake.

**Magnetic Properties**

We have studied the magnetic properties of $Fe_xVSe_2$, for x = 0-0.33, between 2K to 300K and up to 14T fields, the results will be reported in a separate publication [12]. The magnetic susceptibility $\chi$ of all of these, including the compounds of present study follow $\chi = \chi_0 + C/(T - \theta_w)$ dependence around room temperatures. Here $\chi_0$ is the Pauli-paramagnetic contribution (between 3.0 - 4.5 $\times 10^{-4}$ emu/mole) of the electrons in the 3d band of V atoms, C and $\theta$ respectively are the Curie constant and the Weiss temperature related to the paramagnetic moments of the Fe atoms. The Curie constant varies between 0.25 and 1.0 emu-K/mole of Fe atoms in our compounds for $x \leq 0.33$ and corresponds to an effective magnetic moment of 1.4 to 2.8 $\mu_B$/Fe-atom. Hence the Fe atoms are in the low spin state in all of our compounds [11,12]. We have reported the room temperature values of these parameters in table-1, for the two phases of $Fe_{1/3}VSe_2$ namely; FVS700 and FVS550A, which were prepared at $700^0$C and $550^0$C respectively. The low values are in contrast to the Fe moment in the intercalates of Nb- and Ta-dichalcogenides and also to the corresponding compounds of V and Ti, viz. $FeV_2S_4$, $FeV_2Se_4$ and $FeTi_2Se_4$ (with $Cr_3S_4$ structure) [1,6-10]. Moreover, we also find that the values of $\chi_0$ and C in our compounds sensitively depend on the preparation conditions, as can be seen in table 1. At low temperatures the magnetic susceptibility shows characteristic deviations from the high temperature Curie-Weiss dependence. In order to clearly show this behavior we have plotted $(\chi-\chi_0)T$ versus temperature (T) of FVS700 and FVS550A respectively in figure 4 and 5. These plots give the temperature variation of Curie-contribution C ~ $(\chi-\chi_0)T$ , of the localized moments, here we ignore the small value of Weiss constant $\theta_w$ .

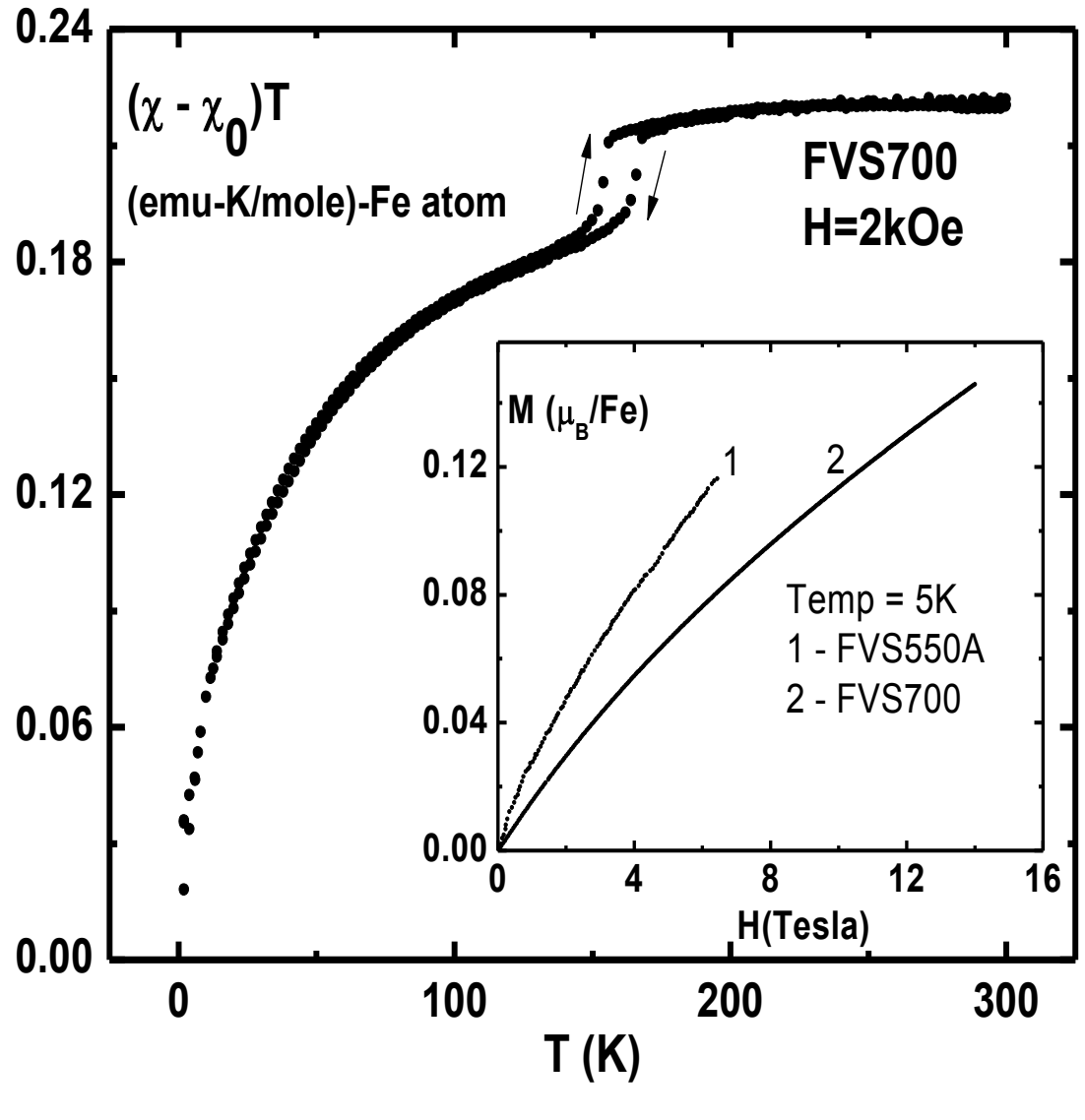

Figure 4. The plots of $(\chi\text{-}\chi_0)T$ giving temperature dependence of the Curie contribution C of fe atoms in FVS700. A rapid reduction of C is seen below 160K transition. M(H) dependence of FVS700 and FVS550A at 5K is shown in the inset.

**Table-1**

| Phase $Fe_{1/3}VSe_2$ | $\chi_0$ ($10^{-4}$ emu/ mole) | C (per mole Fe atom) | $\theta_w$ (K) | $\mu_{eff}$ ($\mu_B$ /Fe –atom) |
|---|---|---|---|---|
| **FVS700** | 4.7 | 0.23 | -5 | 1.35 |
| **FVS550A** | 3.2 | 0.36 | -8 | 1.7 |

The C-value in the high temperature phase of FVS700 is rather small and shows further reduction at 160K-transition. The hysteresis in the magnetic susceptibility around this transition is very similar to the hysteresis in its resistivity as shown in figure 2. The most significant magnetic behaviour is that the C-value reduces continuously in its low temperature phase and attains value of 0.015 emu-K/ mole of Fe atoms, at 2K temperature. Irrespective of the negative value for Weiss constant $\theta_w$, we found no anomalies in the susceptibility that can be related to anti-ferromagnetic interactions.

The results for the FVS550A phase are shown in fig 5. There are significant differences in the susceptibility behavior at high temperatures in this phase. But, the rapid reduction of C ( ~ $(\chi-\chi_0)T$ ) at low temperatures is similar to that found for FVS700 . We also plot the high field magnetisation M measured for both of them in the inset of fig 4. In both the cases M(H) is similar and is consistent with the low value of paramagnetic susceptibility at 2K temperature.

However, there is a marked difference in the susceptibility behavior around 160K-transition for these two compounds as also found in their respective resistance-hysteresis behaviour (see fig 2). We relate these differences to the Fe occupancy resulting from the differently annealed conditions in two compounds, as we have found from our already discussed X-ray diffraction studies.

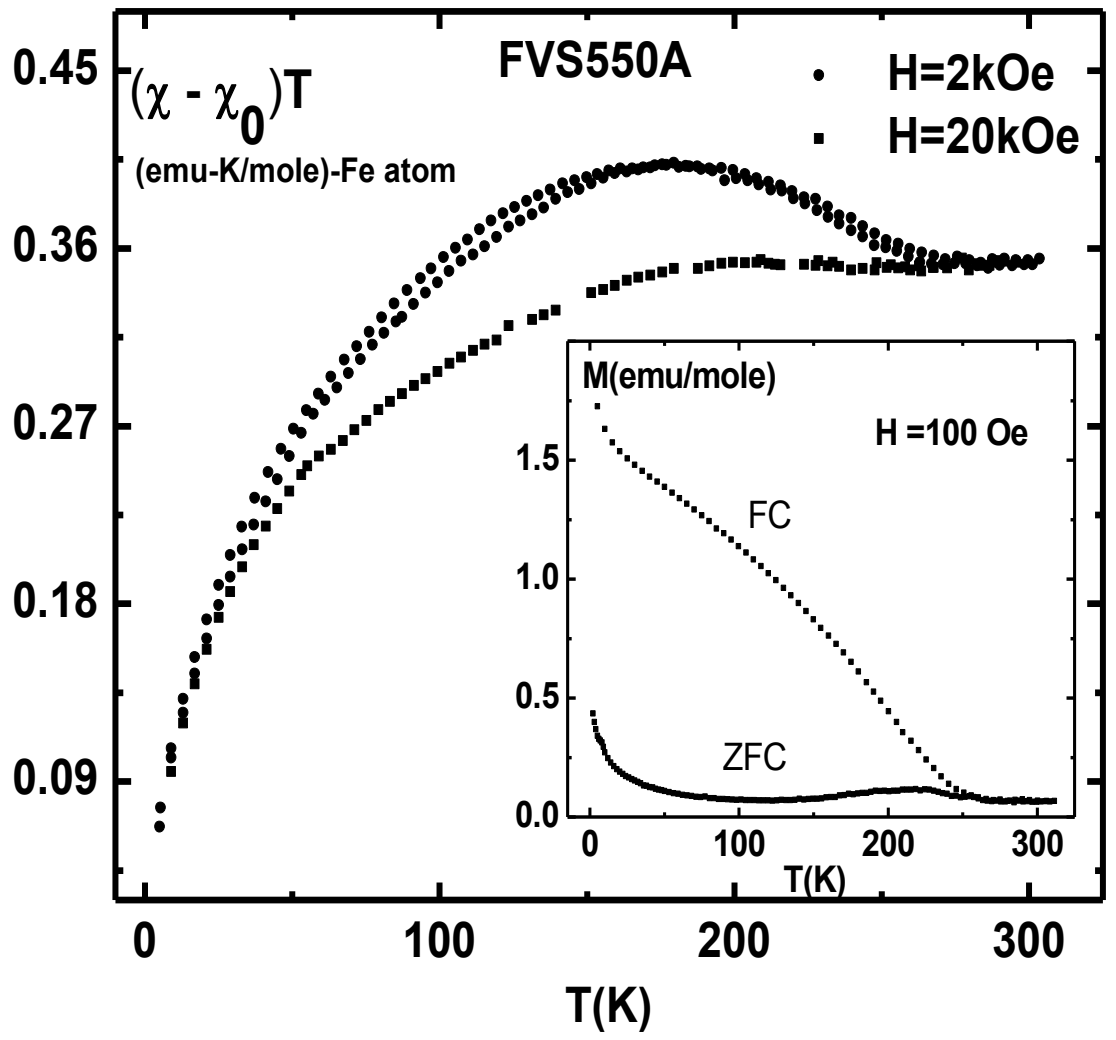

Figure 5. $(\chi-\chi_0)T$ versus temperature dependence of FVS550A found at 2kOe and 20kOe. Between 50K-250K, $\chi$ is field dependent. The field cooled (FC) and zero field cooled (ZFC) magnetisation at 100Oe field is shown in the inset.

At high temperatures, we see a large difference in the Field Cooled (FC) and Zero Field Cooled (ZFC), magnetization behaviour, at 100 Oe field, which starts on cooling the sample below 250K. The FC and ZFC results are plotted in the inset of figure 5. In addition to this irreversibility, we can also see in the main body of fig 5 that the susceptibility is nonlinear between 50K to 250K, since $(\chi-\chi_0)T$ versus T plots depend upon the applied fields of 2kOe and 20kOe. The irreversible and the non-linear magnetisation as shown in figure 5 are normally associated with the inhomogeneous clustering of paramagnetic centers and their freezing in random orientation upon cooling. These properties are consistent with the short range ordering and clustering of Fe atoms as inferred from our X-ray diffraction results of FVS550A.

## Discussion and Conclusion

All the compositions $Fe_xVSe_2$; x = 0 - 0.33 studied by us showed an electronic phase transition on cooling below 160K,with varying degree of the thermal hysteresis in their electrical and magnetic properties [12]. This transition seems to be related to the CDW-instability of the pure phases of 1T-$VSe_2$ at 110K. We find strong precursor effects of the onset of CDW much above 110K in the electrical conductivity of good quality single crystal flakes [12]. In our opinion the intercalation of Fe stabilizes the CDW driven transition at a higher temperature than in pure 1T-$VSe_2$. Our belief can be supported by the Scanning–probe–microscopy studies of super-lattice and CDW related energy gap formation in the Fe-intercalated compounds of 2H-$NbSe_2$ and 2H-$TaSe_2$ [4,5].

The nature of CDW transition at 110K in 1T-$VSe_2$ is very much similar to that in 2H-$NbSe_2$ and 2H-$TaSe_2$ at 35K and 122K respectively. It was found in the STM studies that the up to 33% intercalation of Fe in the latter compounds neither destroys the super-lattice order nor the CDW related energy gap of the host compounds [4]. In our case the CDW instability shifts to a higher temperature on Fe intercalation. The reason for this enhancement of CDW in $Fe_xVSe_2$ may be the reduction in the correlation effects among the conduction electrons of vanadium. The similar effect on the CDW is found by the external pressure in pure 1T–$VSe_2$ [16]. The Fe intercalation entails reduction in c/a, and exerts a positive chemical pressure on the $VSe_2$ layers. The band broadening reduces correlations among electrons, thus strengthening the CDW formation at higher temperatures. We also observe that the short range vacancy ordering in FVS550A specimen, modulate the local structure but does not suppress the CDW formation. The hysteresis of the transition is much wider due to highly strained nature of the crystal structure in this phase.

The magnetic properties of these compounds are found to be very interesting and are quite different from the other 3d intercalates of layered compounds of Nb and Ta [1,6-7], including $FeV_2S_4$, $FeV_2Se_4$ and $FeTi_2Se_4$ compounds [9-10]. In contrast to these our compounds show absence of magnetism at low temperatures. In other compounds , all the intercalate atoms including Fe atoms respect the first of Hund's rule and show magnetic moments with high spin state. In our compounds the Fe moments are unusually low, and correspond to at most only one unpaired electron on each Fe atom. Hence the Fe, which occupies the octahedral positions in van der Waals gap is in the low spin state of $Fe^{3+}$ ions, indicating that the crystal field splitting of d-levels is larger than the Hund's rule coupling energy in these compounds.

Below the CDW transition, we observe a continuous and rapid reduction in the Curie-constant on cooling. There are many possible mechanism for this variation such as, the Van Vleck paramagnetic contribution, the mixed valence nature of Fe ( $Fe^{3+} \leftrightarrow Fe^{2+}$ ) both in low spin state, or the compensation of Fe moment by the conduction electron polarization. Because of the complexity of the crystal structure and in the absence of any knowledge of the local nature of magnetism it is not possible to discus this property. A detailed NMR or Mössbauer resonance study will be very useful to clarify the observed magnetic behavior in our compounds.

## Acknowledgements

We are thankful to A.K. Nigam of TIFR, Mumbai and Alok Banerjee of IUC, Indore for the magnetic measurements. C.S.Yadav acknowledges Council of Scientific and Industrial Research, India for the senior research fellowship grant.